# Maximizing the Value of Enterprise Human-Computer Interaction Standards: Strategies and Applications

Wei Xu

Center for IT Human Factors Engineering, Intel Corporation, USA

## INTRODUCTION

Human factors/ergonomics (HFE) standards are not only a useful reference for experienced HFE practitioners but can also provide guidance for organizations that are inexperienced in HFE design practice. HFE standards can give credibility to the value of introducing user centered methods (Bevan, 2001). As computing technologies advance, knowledge of HFE has spread to the computing related work, and the field of human-computer interaction (HCI) has grown rapidly. Accordingly, HCI standards have evolved for guiding practice. There is great deal of literature concerned with the development and practice of international standards (e.g., International Organization for Standardization/ISO) and national (e.g., The American National Standards Institute/ANSI) HFE or HCI standards, but little on the practice of HCI standards at lower levels, such as at the level of enterprise. This chapter will focus on the practice of HCI standards there. It intends to assess the challenges of enterprise HCI standards from strategy, development, and governance perspectives. Specifically, we discuss the practices at Intel Corporation, a high-tech enterprise environment.

### Hierarchy of HCI standards

Like HFE standards, the HCI standards system is a pyramidal multi-level structure model. From top to bottom, this multi-level model includes international standards, regional/national standards, and enterprise standards.

The highest level in the HCI standards hierarchy is the relevant international standards issued by the ISO Human-System Interaction Committee (ISO Technical Committee/TC 159). The TC 159 has four sub-technical committees that have published 143 standards in the areas of HCI, user experience (UX) and usability, etc. Most of the published standards are related to HCI/UX. For example, in the well-known ISO 9241 series of standards, each includes a sub-series of standards. Within the ISO 9241 series, ISO 9241-210 specifies human-centered design (HCD) methods and principles, the standard, ISO 9241-220, is a standards document for HCD processes (to be published), and ISO 9241-230 are the standards for evaluation methods in HCD (currently being drafted).

The second level in the hierarchy is the regional/national HFE standards, most of which are related to HCI. For example, the U.S. national standards include nearly 40 government standards published by the National Aeronautics and Space Administration (NASA), the Federal Aviation Administration (FAA), etc., and non-government standards published by the Human Factors and Ergonomics Society (HFES), ANSI, etc. Among these, HFES has published two

HCI related standards, including Human Factors Engineering for Computer Workstations (ANSI/HFES 100, 2007), Human Factors Engineering of Software User Interfaces (HFES 200, 2008).

The third level is the HCI standards that are published by a variety of enterprises and organizations, which are called enterprise HCI standards herein. These enterprise standards specify the HCI standards at a detailed level across the industry and organization domains. The standards may be applicable to the external or internal audience of an enterprise. For example, Microsoft has issued detailed user interface (UI) design standards for the Microsoft Windows family of software products. Apple has issued UI design standards for applications across the iOS technology platform. These standards are for external audience, thus ensuring that developers across industries follow the UI design standards when developing specific products. In addition, many corporations, such as Intel, have published a series of HCI standards for internal use. On one hand, these internal HCI standards ensure internal organization and project teams adopt uniform design standards across internal product lines, while on the other hand, these internal HCI standards also enable and facilitate the HCI design practices within a corporation.

Generally speaking, the contents of the HCI standards across the three layers reflect the relationship between inheritance and consistency. The highest level of international standards defines consensus in all aspects, such as, the guiding principles and design principles, rather than the specific design requirements, reflecting the flexibility of specific design within a certain scope. Meanwhile, the lowest level of enterprise standards is more detailed for specific design within a specific domains/platform.

**Value and Challenges of International/National HCI Standards**
As a foundation, the higher-level (international/national) HCI standards do provide benefits. As an example, the approach to software development in the ISO 9241 series is based on detailed guidance and principles for design, rather than precise interface specifications, thus permitting design flexibility while avoiding constraints on design (Bevan, 2001). As a result, these ISO standards often describe principles, not specific solutions for implementation.

Also, these higher-level HCI standards provide potentially complementary approaches to the assurance of usability and UX in solutions during the HCI practice, when being used in combination with lower-level HCI standards such as individual enterprise HCI standards (Bevan, 2001). From a software user interface (UI) design perspective, use of ISO 9241 and ANSI/HFES 100 standards in combination with a style guide based on individual corporation's branding requirements is useful for detailed and consistent branding design on the UI. As a lesson learned, the attempt by the IEEE to develop a standard for drivability of UI (IEEE, 1993) eventually failed to achieve consensus (Bevan, 2001). This approach does not in itself ensure usability.

However, these higher-level HCI standards face challenges in practice. Enterprise HCI standards have been more influential than ISO, and the ISO standards have not been widely adopted (Bevan et al. 2016)**.** The challenges are outlined below.

### *Availability of Best-Known Methods*
To demonstrate the value of the higher-level HCI standards, one must successfully implement these HCI standards in product development within an organization's environment. However, in the HFE and HCI communities, there is little information about how to effectively apply these HCI standards at an organization level (e.g., an enterprise environment) and how to develop own enterprise HCI standards and effectively manage the governance of compliance in practice.

### *Low Perceived Value of HCI Standards*
Influenced by the organizational culture on HFE/HCI, challenges occur because HFE/HCI professionals' expertise and contributions to design can be undervalued (Green, 2002). As a result, HCI standards are difficult to establish and may not be adopted in practice.

### *Challenge of Conformance*
Very few of the higher-level HCI standards specify the UI design precisely, instead defining general principles from which appropriate UI design and procedures can be derived. This gives the standards authority for good professional practice but makes it more difficult to demonstrate conformance (Bevan, 2001).

### *Intended Audience*
As guided by a human-centered design philosophy, there is no perfect set of guidelines in HCI standards: different audiences have different needs (Bevan, 2005). It is difficult for the higher-level HCI standards can be comprehensive enough to serve all purposes of all types of audience across domains. After identifying the intended audience, HCI professionals in enterprises have to specify the scope and the depth of guidelines in the HCI standard to include in a specific domain/platform.

### *Timing Upgrade*
Today, the pace of technological change continues to accelerate, and while all ISO standards are reviewed at least once every five years (Bevan, 2001), higher-level HCI standards can quickly become out of date. This makes it more difficult to assess the conformance of a product to the standards as documented (Reed, et al., 1999). Standards are supposed to contain a single requirement specifying what kind of information shall be provided to demonstrate that the relevant recommendations in the standard have been identified and followed. This approach cannot be guaranteed at these higher-level standards, as so many recommendations are context-specific.

### *Domain-Specific Issues*
It is difficult to find standards relevant to given projects from the myriad domain-specific standards published by many organizations and agencies (Swaminathan & Rantanen, 2014). There are also very few comprehensive standards available for adoption across industry, and this may not even be feasible, as there are so variety of domains across industry.

## The Needs for Enterprise HCI Standards
As discussed above, there are many challenges in implementing international and national HCI standards to meet specific enterprise needs within a corporate environment. Thus, there is a need to develop more specific HCI standards at the enterprise level that are guided by these higher-level HCI standards. Specifically, in our practice at Intel, we recognize that at least two types of HCI standards are needed when delivering both external and internal digital software solutions: design standards and methodological standards.

### *HCI Design Standards*

With the development of new technologies, there have been efforts to develop HCI design standards in the areas of voice, touch, and gestures. However, most of the current standards focus on the design of visual human-machine interfaces, and thus the discussion in this section also focuses on this aspect. Based on our practices over years as a typical use case in the visual UI platform, HCI design standards for enterprises should include the following categories:

- *UI Design Standards:* The standards should adopt the design guidance and principles inherited from related ISO standards and ANSI/HFES-100 standards. In addition, the standards should specifically reflect the enterprise brand requirements for UI design, specific and usable UI design patterns consistently to be used across the enterprise, and the HCI design guidelines across platforms
- *Visual Design Resource Library:* This library should provide a visual design resource based on the enterprise's branding requirements. Thus, projects can quickly reuse the assets for faster and consistent design that will naturally comply with the enterprise UI design standards
- *Conceptual UI Design Pattern Library*: This library should provide various conceptual UI design patterns (i.e., wireframe/low-fidelity prototypes) to guide UI prototyping work across project teams. The purpose is to encourage projects to carry out UI prototyping and reuse appropriate patterns for usable UI design
- *Code-Based UI Component Library*: Driven by both the visual design resource library and the conceptual UI design pattern library, this library should include code-based UI components. A process should be set up to ensure that UI components can be accumulated to the library based on project code work over time. The developers of future projects can directly reuse code-ready UI components, saving development costs and ensuring usable design.

### HCI Methodology Standards

To enable the development of the organizational maturity in HCI, methodological standards are also needed, which also helps adoption of the design standards.

- *User Centered Design (UCD) Process Standards*: The higher-level HCI standards, such as ISO 9241-210/220, can be leveraged to communicate the needs for UCD activities across programs/projects. The enterprise's UCD process standards should define required activities/methods at key stages of the organization's internal product development process and checkpoints for activities from a governance perspective.
- *UX Quality Standards*: The UX quality standards should specify standardized indicators to be used in the organization, the benchmark value of the indicators, the method of verification, the tracking method, and reporting procedures.
- *Organizational UX Maturity Model*: The model should define the levels of maturity, the dimensions (factors) for evaluation of maturity, and specific checklists.
- *Holistic UX Design Standards*: The UI design standards only cover the UI portion. For best HCI practice, a holistic UX approach must be taken, so that additional factors impacting UX can be addressed, including application performance, work flow, UI writing, UI internationalization, user help/training, etc.

## Reasons for Enterprise HCI Standards

### Promoting Consistent Design and Saving Costs

Consistent design means there is no need to set separate design standards for each product. One standard can cover a range of products. From a UX ecological perspective, consistent design means delivering a unified UX across platforms and products. By standardizing design, a central HCI team in an organization (if it exists) can work with development teams to develop UI assets that can be reused across projects (e.g., code-based UI components), helping developers to quickly generate UI, saving time and costs in development, repair, and future upgrades.

### *Improving Product UX*

The detailed UX design specifications defined by enterprise standards are based on previous UCD activities across projects so these design specifications have a guarantee of usability/UX for a family of products. Furthermore, design based on principles of consistency and usability can help the users (for the same product family) to match the mental models they have built using previous products when learning to use new products, thereby reducing the learning time, improving the UX, and reducing user support costs.

### *Promoting Product Brands*

Any product has a certain brand, and its design is crucial for UX and marketing. The design of product brands is often closely related to the UI, so the standardization of UI design is conducive to standardizing the unified product brand design. For example, a page template that reflects the product brand is available across all pages of an enterprise's website.

### *Evangelizing UX Practice within the Enterprise*

The process by which developers refer to UI design standards is an opportunity by itself for them to learn UX. UX practice has been integrated into the standardized UCD development process, which is a very good process for the development team to learn UCD and coordinate with UX personnel. UI design standards and UCD process standards also help the team promote UX, while standardized metrics and benchmarks help the organization's management track UX quality across products.

### *Promoting More Efficient UCD Activities*

UX artifacts, such as a series of personas and user journey maps that are built using standardized user research methods (including user interviews, questionnaires, etc.), can be used to build a cross-project reusable UX artifact repository for reuse across projects in the future. If necessary, new project teams can update these artifacts based on updated user requirements. This will save time for future UCD activities.

### *Helping Outsourcing Contract and Business Acquisition*

Driven by cloud-computing technology and cost control, many enterprises enlist third-party development contractors to complete some product development. Enterprise HCI design standards facilitate the communication of HCI design requirements and acceptance criteria with contractors. In addition, when an organization acquires other companies, a clear established UI design standard is conducive to providing guidance on the HCI design for the products of the acquired company.

### *Facilitating the Growth of the UX Organization*

In a large and medium-sized corporation, the experience of conducting HCI work across internal organizations may be uneven. These organizations need help from the enterprise's central HCI group in methodology and design resources. The development of HCI standards, including design and methodology, will help the dispersed UX personnel adapt to growth. In addition, the process of developing these standards is a learning and growth process for the central group, which is also conducive to enhancing the influence and leadership of the central HCI group in relation to the diverse business product lines.

## DEVELOPING STRATEGY

### UX Strategy

Today, enterprise resource planning (ERP) systems are commonly used in corporations. ERP is a complex computer-based digital environment of solutions that enterprise employees rely on these solutions to do their daily jobs. These solutions include customer relations management (CRM), supply chain management (SCM), human capability management (HCM), finance systems, business intelligence (BI), product support systems, and utility and productivity solutions, amongst others. The huge set of ERP solutions are mixed with home-grown and vendor solutions. Over the years, these were deployed to internal employees with a variety of business and legal requirements, as well as technological requirements. New technology, such as cloud-based SaaS solutions, also adds more vendor solutions. As a result, there can be inefficiency or lack of integration across data, business processes, and UI.

From a UX perspective, enterprise users daily apply a set of solutions to accomplish their tasks. They interact with these solutions through a variety of user touch points, including user awareness/marketing, devices/installation, access/security, solution UI, business process, user support, and user context using data, etc. Therefore, UX is not simply influenced by a single solution or an individual touch point, but instead by how all these solutions and touch points are combined to provide a *unified experience* for end users. If any of these touch points were to break down, it would create a negative experience. Intel is no exception, and a lack of good UX in many individual apps and across apps due to siloed solutions, requires users to use multiple siloed apps to accomplish their job (e.g., SaaS and home grown solutions). Thus, the hybrid environment of today's ERP systems poses a number of serious HCI and UX challenges to the delivery of a unified experience.

Based on extensive research, a unified experience strategy has been proposed to address the UX issues along with corporation-wide digital transformation efforts (Xu, 2014; Xu et al., 2019). To support the unified experience strategy, enterprise HCI standards, processes and an organization maturity model were developed within the IT organization. Below we focus our discussion on enterprise HCI standards.

### Owing Organization of the Enterprise HCI Standards

To support the implementation of the unified experience strategy, the Intel IT Cross-Domain HCI/UX Technical Working Group (TWG) was set up as a horizontal capability and leadership team to drive the HCI standards and governance. The HCI/UX TWG is embedded within the overall IT enterprise architecture and technical governance body.

The TWG's charter are to: (1) drive HCI design strategy; (2) develop and maintain HCI standards; (3) lead and manage the governance of HCI standards. The HCI/UX TWG is chaired by a principle human factors researcher/architect, and meetings are held bi-weekly. The membership includes representatives of HFEs who are working in different business units and also includes representatives from software architect, developer, interaction designer, business architect, and so on. One of the major tasks is to develop/maintain enterprise HCI standards, review/audit the HCI design of cross-platform projects, and approve proposed HCI standards and design assets. Also, the HCI/UX TWG chair serves as representative of the TWG for multiple technical committees, which helps with alignment and evangelization of HCI/UX across organizations.

**Experience of Using Enterprise HCI Standards**

A few years earlier we developed and published a few HCI-related standards, from both design and methodology perspectives. To further understand the gaps, we conducted a UX study with 20 software developers. The data collected were mapped out into the user journey map with major pain points and needs identified (see Figure 1). As shown in Figure 1, there are many pain points along with the developer's end to end experience of using the HCI standards, including "Be aware" (e.g., "I know that there are HCI standards available for us"), "Access" (e.g., "I know where I can find the standards"), "Understand" (e.g., "I understand the content of the standards", "I understand which parts are mandatory vs. recommended"), "Apply" (e.g., "I know how I can apply the standards to my development work"), and "Be governed" (e.g., "I know what governance process we need to follow for compliance"). The major gaps identified and the actions to be taken across the end-to-end experience journey are summarized as follows.

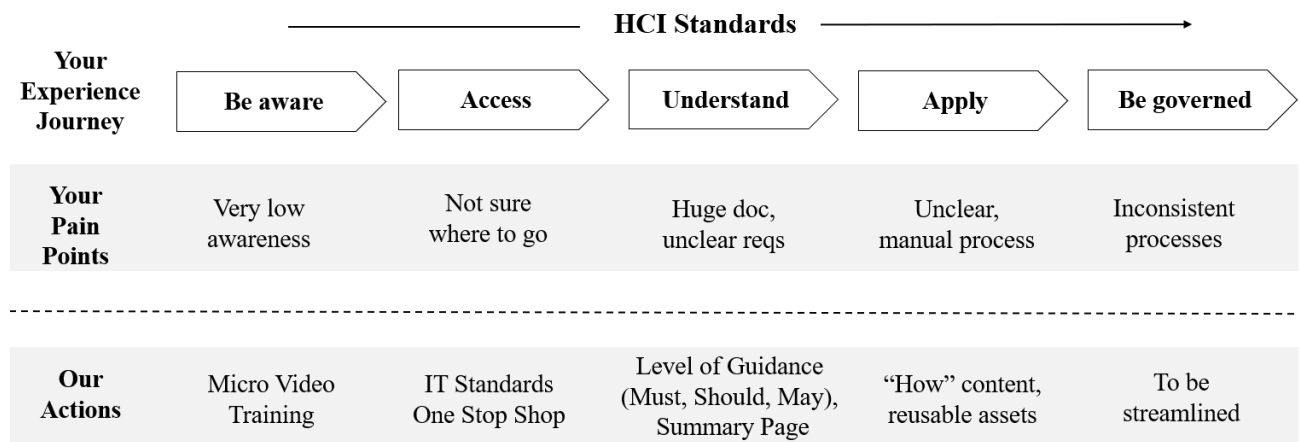

**Figure 1   The gaps identified and actions to be taken**

- *Low Awareness of HCI Standards*: Many people did not know HCI standards or knew only some of them. For the people who knew HCI standards they only knew the basic UI design standards documents (i.e., the look and feel design standards for applications)
- *Not Sure Where to Find HCI Standards*: Many people did not know where to locate these HCI standards or had difficulty finding them
- *Unclear About the Content of HCI standards*: Most people did not fully understand the content of HCI standards documents. In particular, they were not sure whether the statements defined in the standards were requirements (mandatory) or recommendations (optional) from the compliance perspective
- *Not Sure How to Apply the HCI standards:* Lack of reusable assets made it hard for developers to apply the requirements defined in the standards to their development

work
- *Inconsistent Governance Process*: Varied versions of review processes were found across different meetings/forums, and the execution of the review/audit to a large extent relied on the expertise of project technical leaders.

**Developing an Action Plan**

An action plan was developed to specifically address the gaps identified (see Figure 1). The action plan was developed through a process. First, all of the possible actions to address a specific gap were defined, and then prioritized in terms of resources needed and impact. Secondly, the prioritized actions were presented to stakeholders. Finally, the action plan was approved by management and resources were assigned for implementation.

## SOLUTIONS AND EXECUTION

**Transforming HCI Design Assets**

"HCI design assets" broadly refers to HCI related standards, guidelines, tools, and reusable UI resources; that is, anything that positively support HCI work. Based on the strategy and action plan defined, we started a transformation of the existing HCI design assets. Prior to the transformation, there were only a few paper-based HCI standards documents, such as the Master Look and Feel Design Standards for digital solutions, along with some ISO standards (e.g., SIO 9241-100). Table 1 summarizes the "before" and "after" states of various items that were transformed, as well as the benefits obtained.

**Table 1. Transforming UX Assets**

| Before | After | Benefits |
|---|---|---|
| Static documents (Word, PDF, etc.) distributed across places | An online one-stop shop consolidating all documents in digital formats | Easy to access |
| Descriptive guidelines, only covering "do and, "don't" statements | Reusable UI assets, allowing developers/designers to reuse the assets for compliance (e.g., visual, design pattern, UI components) | Reusable, agile |
| A UI-centric approach standards and assets | A UX centric approach by publishing non-UI standards (e.g., app performance, app writing, design guide for persona & user journey map) | Holistic UX approach to minimize UX risk |
| Unclear content of the standards (mandatory vs. recommendation) | Each requirement/statement assigned with RFE labels (must, should, may) | Clarified level of guidance for compliance of standards |

| A "one-way" push model, only allowing audience to access UX assets | A "two-way" model, allowing developers/designers to contribute, rate, comment on UX assets | More interactive and participatory for developers/designers |

### *From Distributed Documents to an Online One-Stop Shop*

In the past, our enterprise HCI standards were distributed to the whole organization in a form of Word/PDF documents, but this approach didn't deliver a good UX as it was hard for users to find and access them. To address this pain point as documented in the user journey map (see Figure 1), we have built an online one-stop shop that consolidates all the HCI standards documents in the form of digital content. An enterprise collaboration (social media) platform is also used to facilitate the participatory approach (see Table 1).

### *From a "Farm of Links" to Searchable, Filterable, and Sortable Content*

Although the one-stop shop had consolidated all the existing HCI standards documents in a central location, it was primarily a farm of links connecting individual standards documents. It was not easy for users to find the content within or across documents. To address the problem, the contents of all the HCI standards documents were converted to a web-based format with meta tags defined across documents to help the implementation of search/filtering/sorting capabilities (see Figure 2). This helped users easily find the content more efficiently and effectively.

| Title | Guidance | Status | Domain | Area | Solution Type | App Layer |
|---|---|---|---|---|---|---|
| Decision Framework For UI Selection | Should Use | Ratified | Application | Development | COTS, SaaS, Cust... | UI |
| HP ALM | Should Not Use | Ratified | Application | Tools | COTS | UI/UX |
| IT Application Internationalization Guidelines | Should Use | Ratified | Application | Development | COTS | SaaS | Cu... | UI/UX |
| IT Application Performance Standards | Must Use | Ratified | Application | Development | | UI/UX |
| IT Axure MLAF Based UI Widget Library | Should Use | Ratified | Application | Development | COTS | SaaS | Cu... | UI/UX |
| IT Master Look And Feel Standards | Must Use | Ratified | Application | Development | COTS | SaaS | Cu... | UI/UX |
| IT Touch UI Design Guidelines | Should Use | Ratified | Application | Development | COTS | SaaS | Cu... | UI/UX |
| IT UI Design Guidelines For Info Visualization, BI Reports, & Dash... | Should Use | Ratified | Application | Development | | UI/UX |
| IT UI Responsive Web Design Guidelines | Should Use | Ratified | Application | Development | COTS | SaaS | Cu... | UI/UX |
| IT UX Governance Standards | Must Use | Ratified | Application | Development | | UI/UX |
| Intel Icon Library | Must Use | Ratified | Application | Development | | UI/UX |
| Intel Writing Style Guide And Editorial Standards For User Interface | Should Use | Ratified | Application | Development | COTS | SaaS | Cu... | UI/UX |
| MEAP Any | Must Not Use | Ratified | Application | Platforms | | |
| Mobile UI Design Standards And Guidelines | Must Use | Ratified | Application | Development | | UI/UX |
| Plugins Native Application | Must Not Use | Ratified | Application | Development | | |
| Plugins Web Browser | Must Not Use | Ratified | Application | Development | | |

**Figure 2. A Screenshot Illustrating the Filterable and Sortable UI**

### *From Descriptive Guidelines to Reusable UI Assets*

The traditional way of delivering UI related HCI standards is based on descriptive guidelines, such as "do" or "don't." For instance, "Don't use x color over x color for low contrast"; "Use x UI design pattern in scenario x." This approach is difficult for developers/designers to follow and it is not efficient, leading to low adoption and difficult compliance from governance perspective. To overcome the issues, we created three reusable UI assets libraries:
- *Visual Design Library*: Includes HTML-based UI style sheets based on the Intel Corporate brand requirements and the enterprise UI design standards (e.g., Bootstraps framework-based); an images/icons library with over 500 icons with different formats per corporate brand requirements
- *Conceptual UI Design Pattern Library*: Includes over 200 UI design concepts (wireframe) in responsive design across desktop, mobile, and tablet platforms. The catalog includes page layout, navigation, various UI elements, and comprehensive page design. All of the design patterns were selected based on usability validation through project work or industry best known methods
- *Code-Based UI Component Library*: Includes many code-based UI components built by UI technology (e.g., Angular JavaScript, React). These UI components are based on some of the conceptual UI design patterns with visual design (style sheet, icons, font, etc.) applied. Thus, developers can easily reuse the UI components during development, which will guarantee usable UI components with compliance to Intel brand requirements by default. Over time, we expect more and more reusable UI components to be added by various project teams.

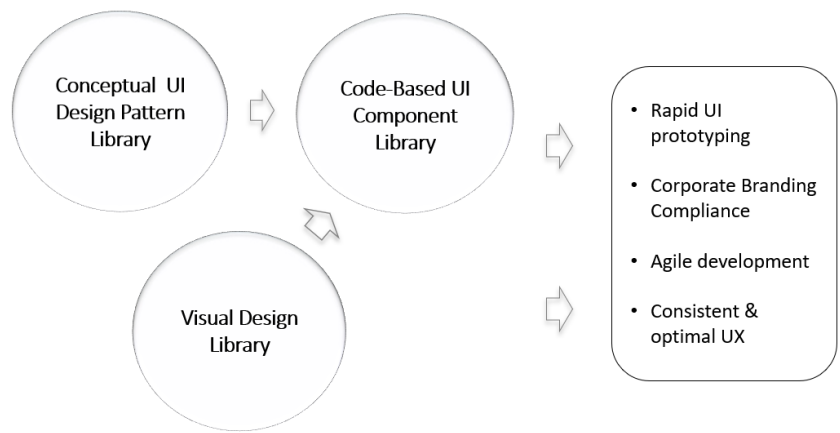

**Figure 3. Usable and Reusable UI assets (the Three Libraries)**

### *From UI-Centric to Holistic UX Approach*
Driven by our unified UX strategy, we have taken a holistic UX approach by developing more UX-centric HCI standards, beyond the originally narrow scope of UI design. Over time, more and more UX-related HCI standards have been (or will be) developed and published. These include:
- *Application Performance Standards:* These specify the pass/fail criteria of the performance (page load/response) test for IT applications
- *Software Accessibility Standards:* These define the requirements of accessibility for software, driven by relevant ISO standards

- *Application Writing Guide:* This provides guidelines for writing on the UI of applications to make content easy to understand, in a consistent style that is necessary to achieve the corporation brand
- *UX Guidelines in Business Process Reengineering (planned):* These guidelines will define the UX guiding principles and methodology that can be applied into the process of redesigning business processes from UX perspective
- *UX Analytics Guide (Support tools):* This guide provides the instructions of using application usage/UX analytics capabilities that help projects to collect users' usage and behavioral data (e.g., click streams) after release. The collected data will help projects identify UX gaps for improvements
- *UX indicators (planned)*: We plan to standardize key UX metrics (e.g., productivity, user satisfaction scores, user support volume/cost) across organizations. The next step is to integrate these UX metrics into the overall operation dashboard of the IT organization, providing higher visibility to management.

### *From Generic UX to Specific UX*

The development of HCI standards should meet the needs of the organization's business and technology strategy. Besides the "generic UX" related HCI standards (e.g., UI design, methods, metrics, etc.), we also develop specific HCI standards as follows:

- *Technology-Driven Approach*: As the pace of technological change continues to accelerate, people are looking for HCI design guidance, and we have published timely and relevant HCI standards, such as UI Design Standards for Mobile Solutions, UX Design Guide for Web Responsive Design
- *Business Strategy–Driven Approach:* Cloud computing and cost control enables a business to seek SaaS/vendor solutions. In order to minimize the UX risk, we published "UX Design Guidelines for SaaS/Vendor Solutions," where key activities are defined, such as, "projects must incorporate UX into the vendor selection scorecard", and "projects must conduct a UX assessment by following a standardized UX assessment template." Some UX/HCI design requirements can also be directly cited in the purchase contracts to help minimize UX risk.

### *From a One-Way Push Model to a Two-Way Participatory Model*

In the past, HCI standards were published in a push model; that is, the HCI organization published standards as Word/PDF-based documents, with the assumption that the standards content would be understandable and easy to adopt by the developers/designers. Over time, we realized that there were problems with this one-way push model, and gradually deployed the following two-way user participatory model. The new approach has helped adoption of the HCI standards and allowed the standards owners to receive feedback for improvement.

- *A User Feedback Mechanism*: A mechanism was set up by leveraging the internal social media platform, where developers/designers can make comments on a standard item or a design pattern, and provide rating scores
- *A User Contribution Mechanism:* For the UI Conceptual Design Pattern Library and the Code-Based UI Components Library, developers and designers can contribute their own assets once validated through project usability activities (e.g., usability testing), and the HCI TWG will send a reward for the contributions. This approach has helped the growth of the libraries and reduced the need for development resources.

**Enhancing Governance of the HCI Standards**

To a large extent, the value of HCI standards will only be realized if they are applied to design and development, which largely relies on whether or not a rigorous governance process is set up. There is industry-wide consensus that a governance model enables an organization to achieve a desired objective (Kiskel, 2011).

### *Level of Compliance*

As shown in Figure 1, one of the typical challenges in governing HFE standards is that either the developers misunderstand the content of the standards (requirements vs. recommendations) or the author of the HFE standards typically fails to clarify the level of guidance (Reeds, et al. 1999). As a result, the developers/designers don't know whether they must or should comply with the standards, making the governance of these standards more difficult. We have adopted the RFC 2119 system, which standardizes the key words for us to indicate requirement levels of standards, to address the challenge (Bradner, 1997), specifically the following tag system that is illustrated in Figure 2:

- *Must Use*: means that the standard/guidance required to be used
- *Must Not Use:* means that the standard/guidance should not be used
- *Should Use:* means that the standard/guidance is recommended, but not enforced
- *Should Not Use:* means that the standard/guidance is not recommended to be used
- *May Use:* means that the standard/guidance is optional.

### *Governance Tracking Process*

One of the success strategies is to follow the requirements defined in the standards throughout the design and development process, don't wait until the end to think about standards (Reeds, et al. 1999). We have integrated the governance of these HCI standards into the existing development process with a tracking process defined. The tracking process of these HCI standards requires projects to be checked or audited at a few key checkpoints along the development process to ensure their conformance to HCI standards. However, in reality, it is not easy to implement all at once. Over time, progress has been made within the organization, from a manual process, followed by a semi-automated tracking process for certain domains, and then transition to a fully automated governance tracking process (which is still not fully realized at this point).

- *Manual Review/Audit Process*: The HCI TWG creates an HCI standards compliance checklist (for the "must use" requirements defined, refer to Figure 2) to be used in Solutions Architecture Review Meetings (weekly). The architecture review meetings are held based on a mature software governance process, which helps facilitate the governance of HCI standards
- *Semi-Automated Tracking Process*: For a few specific domains/platforms (e.g., mobile solutions), a semi-automated workflow-based tracking process is used. This process sets up three domain-specific reviewers for decision-making prior to release, including the domains of security, architecture, and UX/HCI. All mobile projects must activate their accounts of this tracking process at the beginning of development. All three review experts across the three domains must finish their respective reviews, document, and sign off their approval before the project can move on to next steps till release to production. The entire flow is facilitated through email notifications among the three parties
- *Strategic Review*: Besides the tactical reviews as discussed above, the HCI TWG chair attends IT Technical Review Committee (TRC) meetings as a standing member. This is to ensure that major technical decisions across the whole organization will include HCI/UX considerations.

### *Handling Exceptions and the Waiver Process*

In reality, there are always exceptions and waiver requests from project teams. We have set up a sub-process to handle these. Per the process, a project team can request a waiver for compliance of a mandatory requirement specified in these HCI standards by filling out a request form with their supporting reasons and an expiration date for the waiver. The leader of the technical governance committee reviews and dispositions the request (approval or disapproval, and the expiration date if a waiver is granted).

### *Integrating UX Architecture Artifacts into the BDAT Framework*

In the enterprise architecture (software) world, The Open Group Architecture Framework (TOGAF) model has been followed for practices with relatively good governance process implemented (TOGAF, 2019). The TOGAF model defines the architecture and governance across the business, data, applications, and technology (BDAT) domains. Leveraging of the TOGAF model has helped the HCI community integrate HCI architecture artifacts into the BDAT domains.

- *Defining UX Architecture Artifacts:* Key UX architecture artifacts are defined, including the following three:
  - "Experience Gaps and Needs" defines user's pain points when using the "as is" solutions and the needs for the "to be" solutions. The content can be delivered in a format of personas and/or user journey map. The UX architect provides input as requirements into Use Case Diagram, Process Transition Roadmap architects of the business architecture domain
  - "UX prototype" is used as an approach to facilitate projects' iterative activities through UX/UI prototyping, usability testing, and design improvement to minimize the UX risk in solutions
  - "UX Roadmap" defines user's needs over certain period of time in terms of capabilities or usages. We plan to integrate this UX artifact for greater influence on strategy.
- *Integrating UX Artifacts into BDAT Architecture Governance:* Technical architecture governance is mature. For major programs (e.g., digital transformation programs), the UX architecture artifacts are integrated into the Business Architecture domain (i.e., Experience Gaps and Needs, UX Roadmap) and the Application Architecture domain (i.e., UX Prototype).

## Leveraging of An Organization UX Maturity Model

To support the unified UX strategy, we have built an organization UX maturity model with two purposes: (1) a roadmap-based guidance for an organization (program-level) aspiring to a UX-driven organization; (2) an assessment tool of UX capabilities for an organization (program-level) to identify the existing gaps and needs for further steps. Table 2 shows a high-level view of the UX maturity model, detailed requirements have been documented as checklists within each cell of the model. Specifically, UX governance of the HCI standards, along with organization UX strategy, user-centered process, and UX metrics, are included in the model. This is intended to reinforce the implementation of the enterprise HCI standards. There is still more work needed to fully realize the benefits of the UX maturity model.

Table 2   The Summary of the Organization UX Maturity Model

|  | L1: Exploration | L2: In Transition | L3: Sustainable | L4: Proficient | L5: Ideal |
|---|---|---|---|---|---|
| **Organization UX Strategy** | None or limited | Developing | Gaining Commitment | Influencing | Transforming |
| **UX Governance** | Optional & voluntary | Growing | Practicing | Proactive | Collaborative |
| **User-Centered Design Process** | Limited & Inconsistent | Learning | Adopting | Improving | Integrated |
| **UX Metrics** | None or inconsistent | Project-specific | Org. centric & standardized | Continuous improvement | Automated & holistic |

## CONCLUSIONS

In summary, the development of enterprise HCI standards which complement the international/national HCI standards and will maximize the value of the HCI standards across all levels is greatly needed. Specifically, our practices in developing and governing enterprise HCI standards demonstrate that strategic approaches are needed to effectively maximize the influences of enterprise HCI standards, through transformation of HCI design assets, enhanced governance, and the maturation of the UX.

As we move forward, we face the following challenges. First, we are entering the era of artificial intelligence (AI) technology while almost all the HCI standards across international, national, and enterprise levels have been developed specifically for non-AI-based products. There is a need for developing HCI standards in design and methodology, especially at the enterprise level (Xu, 2019).

Secondly, an agile development process has been widely adopted in practice, but the existing HCI standards were developed primarily based on traditional product development lifecycle (e.g., waterfall). Although people claim that the flexible and iterative nature of ISO 9241-210 makes it a good basis for both UX design and an agile development process (Maguire, 2017; ISO, 2010), we do observe the difficulty of integrating HCI process/methods into an agile process, and more best-known methods are needed.

Finally, to further integrate HCI/UX architecture artifacts into the enterprise architecture framework (i.e., the BDAT framework) for maximizing HCI work's benefits, we feel more work is needed, such as, building a solid relationship between UX artifacts and existing key enterprise architecture artifacts, such as business use case diagram, application functional requirement, solution architecture, so that the gaps and needs identified in UX artifacts (e.g., personas) can be traceable and trackable to ensure that they are addressed in the solution.